 \documentclass{emulateapj}

\newcommand{\order[1]}{$~\times 10^{#1}$}

\slugcomment{To appear in ApJ.}

\shorttitle{Search for Gamma Rays from the Centaurus Region}
\shortauthors{Kabuki and Enomoto et al.}

\begin{document}

\title{CANGAROO-III Search for Gamma Rays from Centaurus~A and 
the $\omega$~Centauri Region}

\author{
S.~Kabuki\altaffilmark{1}
R.~Enomoto\altaffilmark{2}
G.~V.~Bicknell\altaffilmark{3}
R.~W.~Clay\altaffilmark{4}
P.~G.~Edwards\altaffilmark{5}
S.~Gunji\altaffilmark{6}
S.~Hara\altaffilmark{7}
T.~Hattori\altaffilmark{8}
S.~Hayashi\altaffilmark{9}
Y.~Higashi\altaffilmark{1}
R.~Inoue\altaffilmark{8}
C.~Itoh\altaffilmark{7}
F.~Kajino\altaffilmark{9}
H.~Katagiri\altaffilmark{10}
A.~Kawachi\altaffilmark{8}
S.~Kawasaki\altaffilmark{2}
T.~Kifune\altaffilmark{2}
R.~Kiuchi\altaffilmark{2}
K.~Konno\altaffilmark{6}
H.~Kubo\altaffilmark{1}
J.~Kushida\altaffilmark{8}
Y.~Matsubara\altaffilmark{11}
T.~Mizukami\altaffilmark{1}
R.~Mizuniwa\altaffilmark{8}
M.~Mori\altaffilmark{2}
H.~Muraishi\altaffilmark{12}
T.~Naito\altaffilmark{13}
T.~Nakamori\altaffilmark{1}
D.~Nishida\altaffilmark{1}
K.~Nishijima\altaffilmark{8}
M.~Ohishi\altaffilmark{2}
Y.~Sakamoto\altaffilmark{8}
V.~Stamatescu\altaffilmark{4}
S.~Suzuki\altaffilmark{14}
T.~Suzuki\altaffilmark{14}
D.~L.~Swaby\altaffilmark{4}
T.~Tanimori\altaffilmark{1}
G.~Thornton\altaffilmark{4}
F.~Tokanai\altaffilmark{6}
K.~Tsuchiya\altaffilmark{1}
S.~Watanabe\altaffilmark{1}
Y.~Yamada\altaffilmark{9}
M.~Yamazaki\altaffilmark{9}
S.~Yanagita\altaffilmark{14}
T.~Yoshida\altaffilmark{14}
T.~Yoshikoshi\altaffilmark{2}
M.~Yuasa\altaffilmark{2}
Y.~Yukawa\altaffilmark{2}
}

\altaffiltext{1}{ Department of Physics, Kyoto University, Sakyo-ku, Kyoto 606-8502, Japan} 
\altaffiltext{2}{ Institute for Cosmic Ray Research, University of Tokyo, Kashiwa, Chiba 277-8582, Japan} 
\altaffiltext{3}{ Research School of Astronomy and Astrophysics, Australian National University, ACT 2611, Australia} 
\altaffiltext{4}{ School of Chemistry and Physics, University of Adelaide, SA 5005, Australia} 
\altaffiltext{5}{ Paul Wild Observatory, CSIRO Australia Telescope National Facility, Narrabri, NSW 2390, Australia} 
\altaffiltext{6}{ Department of Physics, Yamagata University, Yamagata, Yamagata 990-8560, Japan} 
\altaffiltext{7}{ Ibaraki Prefectural University of Health Sciences, Ami, Ibaraki 300-0394, Japan} 
\altaffiltext{8}{ Department of Physics, Tokai University, Hiratsuka, Kanagawa 259-1292, Japan} 
\altaffiltext{9}{ Department of Physics, Konan University, Kobe, Hyogo 658-8501, Japan} 
\altaffiltext{10}{ Department of Physical Science, Hiroshima University, Higashi-Hiroshima, Hiroshima 739-8526, Japan} 
\altaffiltext{11}{ Solar-Terrestrial Environment Laboratory,  Nagoya University, Nagoya, Aichi 464-8602, Japan} 
\altaffiltext{12}{ School of Allied Health Sciences, Kitasato University, Sagamihara, Kanagawa 228-8555, Japan} 
\altaffiltext{13}{ Faculty of Management Information, Yamanashi Gakuin University, Kofu, Yamanashi 400-8575, Japan} 
\altaffiltext{14}{ Faculty of Science, Ibaraki University, Mito, Ibaraki 310-8512, Japan} 

\begin{abstract}

We have observed the giant radio galaxy Centaurus~A 
and the globular cluster $\omega$~Centauri 
in the TeV energy region using the CANGAROO-III stereoscopic system.
The system has been in operation since 2004 with
an array of four Imaging Atmospheric Cherenkov Telescopes (IACT)
with $\sim$100-m spacings.
The observations were carried out in March and April 2004. In total,
approximately 10\,hours data were obtained for each target.
No statistically significant gamma-ray signal has been found
above 420\,GeV over a wide angular region (a one-degree radius from the 
pointing center)
and we derive flux upper limits using all of the field of view.
Implications for the total energy of cosmic rays and the density
of the cold dark matter are considered.

\end{abstract}


\section{Introduction}

Centaurus~A (Cen~A, NGC 5128, J1325$-$4301)
is one of the best examples of a radio-loud AGN.
Viewed at $\sim$60$^{\circ}$ from the jet
axis \citep{graham79,dufour79,jones96},
it has been classified as a ``misaligned" BL Lac 
type AGN~\citep{morganti92}.
Estimates of the distance to Cen~A range from 2 to 8~Mpc.
In this paper, we adopt a value of 3.5\,Mpc~\citep{hui93}.
Due to its proximity and high luminosity,
Cen~A has long been considered 
a good TeV gamma-ray candidate.
A detection of high energy gamma-rays from Cen~A was 
reported in the 1970s,  
however many subsequent attempts have not been successful.

First, the Stellar Interferometer, located near Narrabri, reported 
a positive detection with a flux of 
$ I(>0.3~{\rm TeV})\sim (4.4\pm 1) \times 10^{-11}~
{\rm photon}\,{\rm cm}^{-2}\,{\rm sec}^{-1} $~\citep{grindlay75}. 

At higher energies, the Buckland Park array and the JANZOS Observatory 
also reported the detection of gamma-rays.
The Buckland Park flux, measured between 1984 and 1989, was
$ I(>100 {\rm TeV})\sim (7.4 \pm 2.6) \times 10^{-12}~ 
{\rm photon}$ \citep{clay94}  
and the JANZOS flux, during the period Apr--Jun 1990, was
$ I(>110 {\rm TeV})\sim (5.5 \pm 1.5) \times 10^{-12}~ 
{\rm photon\,{\rm cm}}^{-2}\,{\rm sec}^{-1} $~\citep{allen93}.
Given the significant attenuation expected in the 
gamma-ray flux during interactions with the cosmic microwave background
at these energies, these measured fluxes imply much greater 
intrinsic fluxes 
\citep{pro86}.

\begin{figure}[htbp]
\plotone{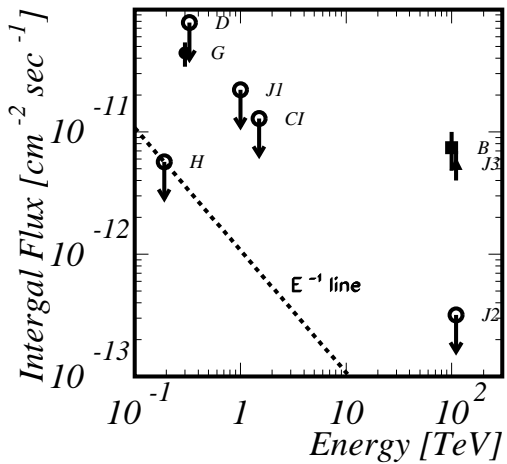}
\caption{
Summary of the past observations. The filled marks are the positive
detection and the open upper limits. The details of these observations
were listed in Table \ref{tintro}. The dashed line is a spectrum proportional
to $E^{-1}$ with H.E.S.S.'s upper-limit level flux. 
}
\label{fintro}
\end{figure}

However, CANGAROO-I~\citep{rowell99}, 
JANZOS (long-term)~\citep{allen93,allen93b},
and Durham~\citep{carraminana} observed the Cen~A nuclear 
region and set upper 
limits on the emission in the VHE range.
Recently, the H.E.S.S.\ group observed Cen~A 
for 4.2~hours in 2004~\citep{aharonian05}. 
Their upper limit was 
5.68 $\times$ 10$^{-12}$ 
${\rm photon}~{\rm cm^{-2}}~{\rm sec^{-1}}$ (1.9\% of the Crab flux)
at 190~GeV.   
They only investigated the region close to the center of Cen~A, 
i.e., the inner jet region.
Cen~A has a large structure
revealed by the radio observations, with inner lobes 
extending over $\sim$10 arcmin, a middle lobe
extending over $\sim$1 degree, and outer lobes 
extending over more than 5 degrees \citep{burns83}.
The fluxes obtained in those observations are plotted in
Fig. \ref{fintro}. The details of the observations are listed
in Table \ref{tintro}.
\begin{table*}[htbp] 
\caption{Details of past observations.}
\label{tintro}
\begin{tabular}{cccccc}
\hline\hline
mark\tablenotemark{a} & reference & epoch & threshold & significance & method \\
\hline
G & \cite{grindlay75} & 1972--1974 & 300 GeV & 4.5 $\sigma$ & Cherenkov\tablenotemark{b} \tablenotemark{c} \\
J3 & \cite{allen93} & Apr-Jun 1990 & 110 TeV & 2\%\tablenotemark{d} & Shower Array \\
J2 & \cite{allen93} & 1987--1992 & 110 TeV & 95\% CL. UL. & Shower Array \\
B & \cite{clay94} & 1984--1989 & 100 TeV & 99.4\% CL. & Shower Array \\
J1 & \cite{allen93b} & 1988 \& 1989 & 1 TeV & 95\% CL. UL. & Charenkov\tablenotemark{b} \\
D & \cite{carraminana} & 1987 \& 1988 & 300 GeV & 3 $\sigma$ UL. & Cherenkov\tablenotemark{b} \\
CI & \cite{rowell99} & Mar--Apr 1995 & 1.5 TeV & 3 $\sigma$ UL. & IACT \\
H & \cite{aharonian05} & Apr 2004 & 190 GeV & 99\% CL. UL. & stereoscopic IACT \\
\hline\hline
\tablenotetext{a}{Mark in Fig. \ref{fintro}.}
\tablenotetext{b}{No imaging Cherenkov telescope.}
\tablenotetext{c}{Spectral index consistent with $\gamma$=-1.7.}
\tablenotetext{d}{Chance probability.}
\end{tabular}
\end{table*}

Cen~A has displayed pronounced variability at X-ray energies over
the last 35 years, with the Grindlay et al.\ detection coinciding
with the peak of the X-ray flux over this period
\citep[see, e.g.,][]{tur97}.
Data from the ROSAT All Sky Monitor \citep{asm},
indicate that in recent years, including the period of the H.E.S.S.\ 
observations, Cen~A has been near its historical minimum X-ray state
(MJD from 53111 to 53113).

CANGAROO-III has data outside of this minimum period
(MJD from 530080 to 530107). 
We also report the results over a wide region.
Part of the southern outer lobe of Cen~A was located within the field of
observations made of $\omega$~Centauri ($\omega$~Cen).
We, therefore, include the results of the latter observation
in this report.
Located at a distance of 4.9~kpc, $\omega$~Cen
is one of the oldest, heaviest globular clusters in our
Galaxy. As globular clusters have been found to contain 
large numbers of millisecond pulsars, they are interesting objects
for high-energy gamma ray observations.

\section{CANGAROO-III Stereoscopic System}

The use of imaging atmospheric Cherenkov telescopes (IACTs) was
established with the statistically unassailable
detection of the Crab nebula at TeV energies by the
Whipple group \citep{whipple}.
This technique enables TeV gamma-rays to be selected from the
huge background of cosmic rays with the use of the 
``image moments" of the shower images \citep{hillas}.
Stereoscopic observations, which allow the signal-to-noise ratio
to be significantly improved, were pioneered by the HEGRA group \citep{hegra}.
The H.E.S.S.\ group has recently reported the detection of 
faint gamma-ray sources with an angular
resolution as fine as a few arc-minutes \citep{HESS_science}.

CANGAROO-III is one of two major IACTs located in the southern
hemisphere.
The CANGAROO-III stereoscopic system consists of four imaging atmospheric
Cherenkov telescopes located near Woomera, South Australia (31$^\circ$S,
137$^\circ$E).
Each telescope has a 10\,m diameter segmented reflector, 
consisting of 114 spherical mirrors 
made of FRP \citep{kawachi}, each of 80\,cm diameter,
mounted on a parabolic
frame with a focal length of 8\,m.
The total light collecting area is 57.3\,m$^2$.
The first telescope, T1, which was the CANGAROO-II telescope,
is not presently in use due to its smaller field of view
and higher energy threshold.
The second, third, and fourth telescopes (T2, T3, and T4) were used for the
observations described here.
The camera systems for T2, T3, and T4 are identical and their details
are given in \citet{kabuki}.
The telescopes are located at the 
east (T1), west (T2), south (T3) and north (T4)
corners of a diamond 
with sides of $\sim$100\,m \citep{enomoto_app}.

\section{Observations}

The observations were carried out 
in the period from 2004 March 16 to April 19
using ``wobble mode"
in which the pointing position of each telescope was
shifted in declination between $\pm$0.5 degree 
every 20 minutes \citep{wobble}
from each target:
(RA, dec [J2000]) = (201.365$^\circ$, $-$43.019$^\circ$) for Cen~A,  
(RA, dec [J2000]) = (201.691$^\circ$, $-$47.477$^\circ$) for $\omega$~Cen. 
One of the reason why we took "wobble" observation is to enlarge
the effective FOV, the other is to average the responses of
individual pixels.
We, therefore, took LONG OFF source run of ''wobble" mode
for background subtractions in the later analysis.

Data with GPS time stamps were recorded for T2, T3 and T4 individually when
more than four photomultiplier (PMT) signals 
exceeded 7.6 photoelectrons (p.e.). 
In the offline analysis stage we combine all these data when
the three telescope's GPS times coincide.
The typical trigger rate was 11\,Hz for three-fold coincidences.
Each night was divided into two or three periods, i.e., ON--OFF,
OFF--ON--OFF, or OFF--ON observations. ON-source observations were timed
to contain the meridian passage of the target. 
On average the OFF source regions were located with an offset in RA of 
+30$^\circ$ or $-$30$^\circ$ from
the target. The OFF-source observations were also made in
wobble mode.
One day was dedicated to ON and OFF source observations of one target,
with the following day dedicated to the other target.
The images in all three telescopes were required to have clusters
of at least five adjacent pixels exceeding a 5\,p.e.\ threshold
(three-fold coincidence).
The event rate was reduced to $\sim$7.5\,Hz by this criterion.
Looking at the time dependence of these rates, we can remove data
taken in cloudy
conditions. This procedure is the same as the ``cloud cut''
used in the CANGAROO-II analysis \citep{enomoto_nature}.
The effective observation times 
for ON and OFF source observations were
639.4 and 586.9~min for Cen~A, and 
600.8 and 429.4 min for $\omega$~Cen.
The mean zenith angles were 17.4$^\circ$ 
and 20.6$^\circ$, 
respectively.

The light collecting efficiencies, including the reflectivity
of the segmented mirrors, the light guides, and the quantum efficiencies
of photomultiplier tubes were monitored by a muon-ring analysis
\citep{enomoto_vela} with the individual trigger data in the
same periods. 
The light yield per unit arc-length is approximately proportional
to the light collecting efficiencies.
The average ratios of these at the observation period with respect to the
mirror production times (indicating the amount of deterioration) 
were estimated to be 70, 70, and 80\% for
T2, T3, and T4, respectively. The measurement errors are considered to
be at less than the 5\% level.
Deterioration is mostly due to dirt and dust settling on the
mirrors.

\section{Analysis}

Most of the analysis procedures used are identical with those 
described in \citet{enomoto_0852}.
As a full instrumental description was given in
\citet{enomoto_vela}, we omit a detailed discussion here.
There are some improvements in the analysis procedure
from that in the previous paper and so we concentrate
on those points here.

At first, the image moments \citep{hillas} 
were calculated for the three telescopes' images.
The gamma-ray's incident directions were determined by minimizing
the sum of squared widths ($\chi^2_0$: weighted by the photon yield) 
of the three images seen from the assumed position (fitting parameter)
with a constraint on the distances from the intersection point to each
image center ($D_{IP}$).
The $D_{IP}$ can be estimated from the ratio $Length/Width$.
The prediction curve ($f$) and its error ($\sigma)$ 
were estimated using Monte-Carlo simulations.
The constraint is;
$$\chi^2_c=\frac{[D_{IP}-f(Length/Width)]^2}{\sigma^2}.$$
In order to balance dimensions between $\chi^2_0$ and $\chi^2_c$, we
need to multiply $n\cdot <Width^2>$ by $\chi^2_c$, where $n$ has the units of
number of photo-electrons (p.e.) and is optimized 
by Monte-Carlo simulations ($n$=10 p.e.\ for this analysis).
Finally the minimizing variable is;
$$\chi^2~=~\chi^2_0~+~n\cdot <Width^2>\cdot \chi^2_c.$$
For large images, the former term dominates and for small images
(for example, events at large zenith angles),
the latter dominates. With this fit, we can improve uniformity
of $\theta^2$ resolution especially with respect to zenith angle.

In order to derive the gamma-ray likeliness,
we used  
the Fisher Discriminant (hereafter $FD$) \citep{fisher,enomoto_vela}.
Input parameters were
$$\vec{P}=(W2,W3,W4,L2,L3,L4),$$
where $W2,W3,W4,L2,L3,L4$ are energy corrected $Widths$ and $Lengths$ for the
T2, T3, and T4.
$FD$ has a small dependence on the zenith angle, i.e., as the zenith
becomes larger, images become smaller ($FD$ larger).
We corrected for this using Monte-Carlo simulations.

We rejected events with any hits in the outermost layer of the cameras
(``edge cut"). These rejected events cause finite deformations especially
in the $Length$ distribution which results in deformations of the $FD$. 
In this analysis, we allowed less energetic hit pixels on the "edge" layer
if their pulse heights were less than that of highest 15 pixels. 
This cut was improved in order to increase acceptance.

Since we have $FD$ distributions for OFF-source data and the 
Monte-Carlo gamma-ray
events, we can assume these are background and signal events respectively.
Note that in the gamma-ray simulations we used a spectrum 
proportional to $E^\gamma$ where $\gamma$=$-$2.1.
We can therefore fit the $FD$ distribution of ON
with the above emulated
signal and real background functions, to derive the number of signal events
passing the selection criteria. 
With this fit, we can determine the gamma-ray excess without any positional
subtractions.
This is a one-parameter fitting with the constraint that
sum of signal and background events corresponds to the total number of events.
These coefficients can be derived exactly analytically.

\section{Results}

\begin{figure}[htbp]
\plotone{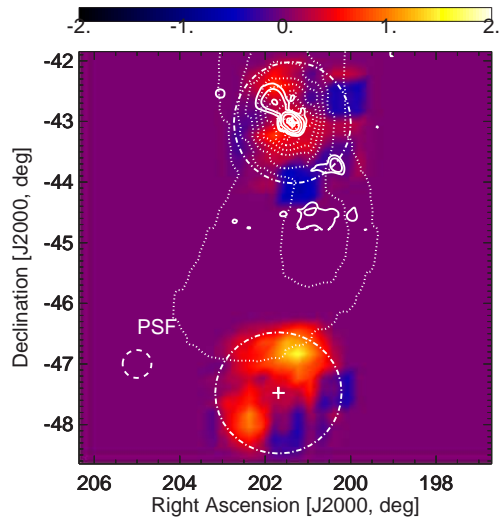}
\caption{Significance map. 
The angular bins are 0.2\,$\times$\,0.2 degree$^2$
squares. This was obtained by the fitting procedure described in the
text. For the background function of the fit, the OFF region of a 
0.6\,$\times$\,0.6 deg$^2$ 
square centered on (but not including) the corresponding bin was used.
The smoothing was carried out averaging the neighboring bins.
The crosses are the pointing centers.
Events within one degree radius circles are plotted (the dot-dashed circle).
The point spread function (PSF) is shown by the dashed circle. 
The solid contours are 4850\,MHz radio data of
the inner lobes and middle lobe and the dotted contours are 408\,MHz 
observations of the outer lobes. 
}
\label{add}
\end{figure}

The morphology (significance map) is shown in Fig.~\ref{add}.

The FOV was binned to 0.2 degree\,$\times$\,0.2 degree squares.
The $FD$ distribution inside each bin was made. The signal
function (for gamma rays) was made using the Monte-Carlo simulation.
The background (for protons) function was made collecting OFF events
inside 0.6\,$\times$\,0.6 degree$^2$ centered around the corresponding
signal bin.
The one-parameter $\chi^2$ fit was carried out and the excess counts
were obtained inside a one-degree circle (the dot-dashed circle) 
centered at the average pointing position.
The pointing centers are shown by the crosses. At one degree radius, the
acceptance decreases by 30\%.
There is, therefore, not any statistically significant excess anywhere.
The positive and negative fluctuations roughly agree
in the Cen~A field of view. In that of $\omega$~Cen
a positive offset was observed ($\sim$0.4 $\sigma$), however, it is
not statistically significant.
The peak located at north of $\omega$-Cen field and its significance
is 1.5 $\sigma$.
The point spread function (PSF) is shown by the dashed circle
at left-lower place.
The solid contours are obtained via \cite{skyview} and are the 
GB6 radio data (4850\,MHz) 
showing the inner lobes and middle lobe, 
and the
dotted contours are 408\,MHz radio data, roughly 
showing the outer lobes of Cen~A from Fig.\,11 of \cite{burns83}.

In order to derive a UL for a point source, we choose
$\theta^2$=0.06 degree$^2$ from the center of Cen A  for the cut position.
The 2$\sigma$ UL was obtained to be 39.8 events.
The energy threshold for this analysis was obtained to be 424\,GeV
by the Monte-Carlo simulation.
For $\omega$-Cen,
a 2$\sigma$ UL of 32.3 events was obtained.

\begin{figure}[htbp]
\plotone{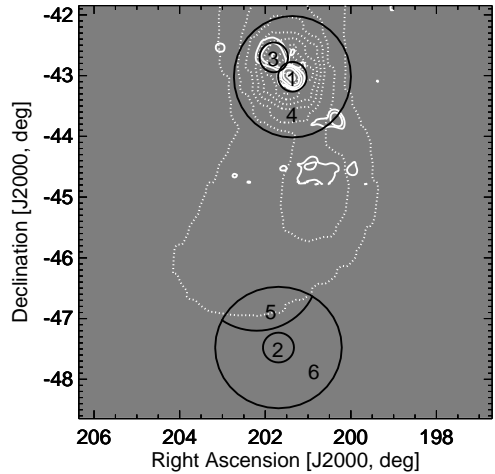}
\caption{Definition of search regions.
Region~1 contains the jet and inner lobes of Cen~A 
and region~2 is $\omega$~Cen.
These sizes are identical to the PSF.
Region~3 is the middle lobe of Cen~A, centered at 
(RA, dec [J2000]) = (201$^\circ$.95, $-$42$^\circ$.85). 
The size is identical to the PSF.
Region~4 is within a one-degree circle from the center of Cen~A
not including regions 1 and 3: which contains portions of 
the outer lobes of Cen~A close to the nucleus.
Region~5 is the overlap of two one-degree circles centered
at (RA, dec)=(205$^\circ$, $-$47$^\circ$) and $\omega$~Cen: the 
southern extremity of the outer lobe
of Cen~A \citep{burns83}.
Region~6 is within one-degree circle from the center of $\omega$~Cen
not including regions 2 and 5.
}
\label{region}
\end{figure}

In order to derive flux ULs for the jet and lobes
of Cen~A, and $\omega$~Cen, we defined the search regions
as shown in Fig.~\ref{region}.
Region~1 contains the jet and inner lobes of Cen~A, region~3 
contains the middle lobe, and 
regions~4 and 5 contain portions of the outer lobes, 
while region~2 contains $\omega$~Cen.
There is no high-energy astronomical counterpart in region~6
(although low level radio emission from the southern outer lobe of Cen~A 
may extend into this region).
The flux upper limits for various regions were obtained from the
$FD$ distributions of the ON-source runs for corresponding areas.
The OFF-source runs were used for the background $FD$ functions
and the Monte-Carlo simulations for gamma rays for the signal.
We do not observe any statistically significant excesses
in any regions. 

Thus we can obtain flux ULs for region~1.
Those of Cen~A center region are summarized in Table~\ref{table_1}.
\begin{table}[htbp]
\caption{Flux Upper Limits above various energy thresholds for the Cen~A jet}
\label{table_1}
\begin{tabular}{ccc}
\hline\hline
Excess Upper Limit &Energy Threshold & Flux Upper Limit \\
 Events          &GeV            & ${\rm cm}^{^2}{\rm s}^{-1}$\\
\hline
39.8 &  ~424 &  0.491\order[-11] \\
12.6 &  2074 &  0.903\order[-12] \\
~5.0 &  6202 &  0.485\order[-12] \\
\hline\hline
\end{tabular}
\end{table}
ULs are presented for three different energy thresholds.

Exactly the same procedure was carried out for the $\omega$~Cen
region. 
Again there is no statistically significant excesses.
The flux ULs for region~2 are summarized in Table~\ref{table_2}.
\begin{table}[htbp]
\caption{Flux Upper Limits above various energy thresholds for $\omega$~Cen}
\label{table_2}
\begin{tabular}{ccc}
\hline\hline
Excess Upper Limit &Energy Threshold & Flux Upper Limit \\
 Events          &GeV            & ${\rm cm}^{^2}{\rm s}^{-1}$\\
\hline
32.3 &  ~471 &  0.355\order[-11] \\
~7.9 &  2216 &  0.519\order[-12] \\
~7.0 &  6518 &  0.602\order[-12] \\
\hline\hline
\end{tabular}
\end{table}
The flux ULs for the regions~3, 4, 5, and 6 are summarized in 
Table~\ref{table_other}.
\begin{table}[htbp]
\caption{Flux Upper Limits for the Other Regions}
\label{table_other}
\begin{tabular}{cccc}
\hline\hline
Region&Excess Upper Limit &Energy Threshold & Flux Upper Limit \\
& Events          &GeV          & ${\rm cm}^{^2}{\rm s}^{-1}{\rm Sr}^{-1}$\\
\hline
3& ~37.3&   424&  0.875\order[-7]\\
4& 132.6&   424&  0.193\order[-7]\\
5& 203.7&   471&  0.106\order[-6]\\
6& 304.3&   471&  0.436\order[-7]\\
\hline\hline
\end{tabular}
\end{table}
Here, the ULs are divided by the solid angles 
of the observations,
because the inner lobes, middle lobe, and outer lobes are candidates for
diffuse emission.

\section{Discussion}

First, our flux ULs together with that of H.E.S.S.\ 
are quite low compared to the past indications of TeV
gamma rays and also to the past UL measurements,
demonstrating the strength of the stereo-IACT technique.

Next we discuss the total cosmic-ray energy and the cold dark matter (CDM)
density for each corresponding astronomical object.
Here, we assumed distances of Cen~A and $\omega$~Cen to be
3.5\,Mpc and 6\,kpc, respectively.
For the total energy of the cosmic rays, we assume that the
origin of the TeV gamma-rays is the inverse Compton (IC) scattering of
cosmic-ray electrons with the cosmic microwave background (CMB).
The electron/proton ratio of our Galaxy is thought to be 0.1$\sim$1\%.
Therefore, 10$^{52}$\,erg for the electron cosmic-ray component might be a 
standard value for typical galaxies.
Here, we assumed the electron energy spectrum of $\propto E^{-2.1}
e^{-\frac{E}{E_{max}}}$.
We integrate the electron spectrum greater than 1\,GeV
in order to calculate the total energy of electron component.
The interstellar matter densities are not well measured in the
corresponding regions. The electron assumption is reasonable,
because the CMB density is well defined and can be considered to
a lower limit for ambient photon density.

\begin{figure}[htbp]
\plotone{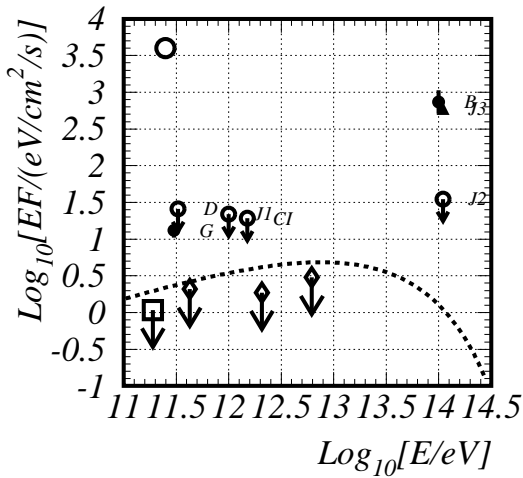}
\caption{
SED for the jet region of Cen~A.
The open circle is the prediction of \cite{bai01}.
The black square with the arrow is the H.E.S.S.\ upper
limit \citep{aharonian05}.
The open diamonds are our upper limits.
The expected gamma-ray yield with a total electron energy of
10$^{54}$\,erg with a 100\,TeV cutoff energy is shown by the dashed
curve.
The details of the marked observations
were listed in Table \ref{tintro}.
}
\label{sed_a}
\end{figure}

Cen~A is considered to be a low-energy peaked BL~Lac object (LBL) rather than
high-energy peaked object (HBL).
The spectral energy distribution (SED) is plotted in Fig.~\ref{sed_a}.
Even if a cosmic ray flux similar to our Galaxy's exists, 
there is no contradiction with any past measurements in the all wave
length region \citep{ned}.
This, therefore, is not denying higher energy component such as 100~TeV
even in the higher level compared to our Galaxy, for example 10$^{53}$\,erg
of electrons.
Actually, \cite{bai01} predicted a 
time-dependent huge energy flow in the TeV energy region
which is shown by the open circle in Fig.~\ref{sed_a}.
The H.E.S.S.\ UL is shown by the open square with arrow
and our ULs by the open diamonds.
These are orders of magnitude lower than the predicted value.
However, as our measurement periods correspond to a quiet phase of
Cen~A (from ASM data), there is no direct contradiction of this idea.

\begin{figure}[htbp]
\plotone{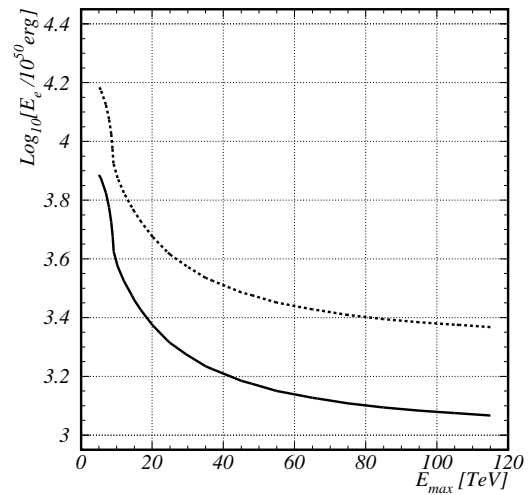}
\caption{One- and two-$\sigma$ upper limits to the total energy of
the electron cosmic ray flux inside the jet region of Cen~A
versus the maximum accelerated energy.
The one-$\sigma$ line is the solid curve and the two-$\sigma$ line is
dotted.}
\label{ee_a}
\end{figure}

We rather concentrate ourselves to the quiet, stable, and average states
of Cen~A.
If a high energy component of electrons $\propto~E^{-2.1}e^{-\frac{E}
{100 {\rm TeV}}}$
exists at the level of the total energy of 10$^{54}$ erg which is
one-hundred times stronger than that of our galaxy,
the expected TeV gamma-ray flux is the dashed curve in Fig.~\ref{sed_a}.
That is higher than our ULs, i.e, we can derive a meaningful UL
to the total electron densities.
The 2$\sigma$ ULs (the dotted curve) for the above assumption were derived
and shown in Fig.~\ref{ee_a}.
The vertical scale is the logarithm of total electronic energy
normalized to 10$^{50}$~erg and the horizontal scale is the
maximum accelerated energy of the electron in TeV.
At 100\,TeV, the limit is 3$\times$10$^{53}$\,erg
which is thirty times bigger than that of our Galaxy.
Although Cen~A is highly active especially at radio,
IR, and MeV energies compared to our Galaxy, 
any high energy cosmic-ray components are not detected;
Cen~A is a mysterious object.
If we assume that the target photons are of those IR temperature
blackbody radiation, the limits of the total energy
will be significantly reduced.
The future Cherenkov telescope array experiment (CTA) \citep{cta}
is promising to help resolve this, as it aims
to be two orders of magnitude more sensitive than
the present IACTs.
If CTA does not detect any signal from this region, we might
need to totally revise the idea of the cosmic-ray origin.

According to the ASM data \citep{asm}, the X-ray activity of Cen~A
was at a minimum in the MJD range between 53111--53113.
The H.E.S.S.\ observation was completely coincident with this period,
and ours partially.
We, therefore, can divide the observational period into two,
i.e., coincident with the ASM minimum, and the remaining data.
The ULs for these two periods are summarized in Table \ref{table_time}.
\begin{table}[htbp]
\caption{Flux Upper Limits in two epochs for Cen~A jet}
\label{table_time}
\begin{tabular}{cccc}
\hline\hline
Excess Limit &Energy Threshold & Flux Limit & MJD-53000 \\
 Events          &GeV          & ${\rm cm}^{^2}{\rm s}^{-1}$ & day\\
\hline
57.6&   424& 0.128\order[-10]& 80, 82, 89, 91, 107\\
26.7&   424& 0.737\order[-11]& 111, 113\\
\hline\hline
\end{tabular}
\end{table}
There is however no excess in the period of
the (slightly) higher state.

$\omega$~Cen is the biggest and oldest globular cluster inside our
Galaxy.
There is no ionizing gas which is considered to be an origin of
cosmic rays.
We, therefore, do not expect any TeV radiation from this object.
However, to know the minimum level of cosmic-ray density is important.
There may be a lot of milli-second pulsars as has been found in other
globular clusters.
The object may still hold a CDM halo \citep{peebles},
as these are not clearly denied at the present level of measurements.
The ULs for the total energy of electron cosmic ray component is derived
in the same way as described previously.
Under the assumption of a maximum electron accelerated energy of 100\,TeV,
the total electron energy is less than 4 $\times$ 10$^{47}$ erg, i.e., 
less than one-supernova level, a surprisingly quiet object 
considering its total mass of $\sim 5\times$10$^6$\,M$_\odot$
\citep{vandeven, meylan, richer}.

Cen~A is a massive object and $\omega$~Cen a much higher density object
than our Galaxy as a whole.
The upper limit study of the CDM density would provide
meaningful information as to the mechanism of their formations.
However note that for the case of globular clusters, CDM was considered
to be stripped off via tidal force in the periodic motions 
inside the parasite galaxy.
\cite{peebles} discussed the existence of the dark matter
in this kind of object.
The CDM search was carried out following the procedure of \cite{enomoto_cdm}.
Here we additionally used the EGRET UL at 189 MeV \citep{ned}.
This is effective around the sub-TeV CDM particle mass region.
Assuming the CDM particle mass of 1 TeV, the 2-$\sigma$ upper limit
for the CDM density is 1.2 $M_\odot$/pc$^3$.
These are quite high compared to our local density estimation of
CDM of 0.01M$_\odot$/pc$^3$.
The mass of Cen~A was estimated to be several $\times$10$^{11}$M$_\odot$
\citep{mathieu, peng1, peng2, karachentsev, woodley}.
In order to derive the total mass of the CDM, we multiply by a factor of
1.42 $\times$ 10$^{13}$~pc$^3$ considering our angular resolution
of $\theta^2$=0.06~degree$^2$ with an assumption of the distance
of 3.5~Mpc \citep{hui93}.
Then the total CDM mass inside the volume is less than 2 $\times$
10$^{13}$ M$_\odot$, i.e., a meaningless result.

\begin{figure}[htbp]
\plotone{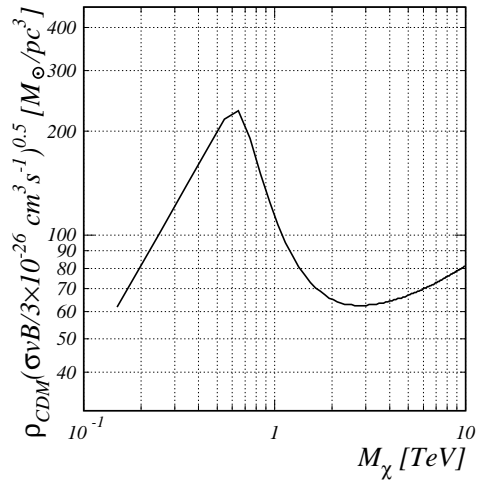}
\caption{Upper limits of the CDM density versus particle
mass assumption for $\omega$~Cen.}
\label{cdm_omega}
\end{figure}

For $\omega$~Cen, we can carry out the same discussion
and the results are shown in Fig.~\ref{cdm_omega}.
The result is one order of magnitude higher than that of Cen~A.
This is due to distance of 6\,kpc compared to that of Cen~A of 3.5\,Mpc
while the volumes are determined by the same angular resolution
of the observation.
However, multiplying the volume factor of 4$\times$10$^{4}$~pc$^3$,
then the most CDM-particle-mass region is under its total gravitational 
mass of 5$\times$10$^6$~M$_\odot$.
In most regions of CDM particle mass between 100\,GeV and 10\,TeV,
we can reject the hypothesis that the gravitational mass is dominantly
occupied by the CDM particles, they might be absorbed by our Galaxy.
Presently, high energy physicists believe that the CDM particle mass
lies inside the above region.
The future CTA project \citep{cta} would reveal the fact in a wider mass
region, or the near-future LHC project might give a conclusive result
on this subject.
In cosmological life,
the larger eats the smaller but its core remains.

\section{Conclusion}

In this paper, we present the results of stereoscopic observations 
of Centaurus~A and $\omega$~Centauri 
with the CANGAROO-III telescopes.  
The observation period was 2004 March 16 to April 19 and 
the total observation times were 640 min for Cen~A and 600 min for
$\omega$~Cen.
The observation of Cen~A was carried out over a similar period
to that of H.E.S.S.
The $\omega$~Cen observations are the first reported
trial by Imaging Atmospheric Cherenkov Telescopes.
The analysis was carried out inside a one-degree (radius) circle from
the average pointing position.
We derived flux upper limits for regions containing
the jet and inner lobes, the middle lobe, and portions of the
outer lobes of Cen~A, and center of $\omega$~Cen.
The Cen~A upper limits are,
as with the H.E.S.S.\ limit, an order of magnitude lower than 
previous measurements.

The upper limits for the total energy of the electron components
of cosmic rays were calculated under the assumption of Inverse
Compton Scatterings on the cosmic microwave background, with limits of
3$\times$10$^{53}$ and 4$\times$10$^{47}$~erg obtained for Cen~A and
$\omega$~Cen, respectively.

Finally we gave upper limits to the density of Cold Dark Matter (CDM). 
Around the TeV region, we obtained upper limits
of its density of 2~M$_\odot$pc$^{-3}$ for Cen~A and
100~M$_\odot$pc$^{-3}$ for $\omega$~Cen.
The limit for Cen~A was greater than its gravitational mass,
however, that for $\omega$~Cen was less than it.

\acknowledgments

This work was supported by a Grant-in-Aid for Scientific Research by
the Japan Ministry of Education, Culture, Sports, Science and Technology, 
the Australian Research Council, JSPS Research Fellowships,
and Inter-University Researches Program 
by the Institute for Cosmic Ray Research.
We thank the Defense Support Center Woomera and BAE Systems.

\end{document}